\documentstyle[preprint,revtex]{aps} 
\begin{document}
\draft
\preprint{ETH, Lausanne}
\preprint{April 1996}
\begin{title}
Ferromagnetism in an itinerant electron system:\\
Hubbard model on complete graph
\end{title}
\author{D. F. Wang}
\begin{instit} 
Institut de Physique Th\'eorique\\
Ecole Polytechnique F\'ed\'erale de Lausanne\\
PHB-Ecublens, CH-1015 Lausanne, Switzerland
\end{instit}
\begin{abstract}
In this work, the ground states of the Hubbard model on 
complete graph are studied, for a finite lattice size $L$
and arbitrary on-site energy $U$. We construct explicitly
the ground states of the system when the number of the electrons
$N_e\ge L+1$. In particular, for $N_e=L+1$, the ground state is 
ferromagnetic with total spin $s_g=(N_e-2)/2$. 
\end{abstract}

\narrowtext

Hubbard model has been of considerable interests since the discovery
of the high temperature superconductivity\cite{anderson}. 
In one dimension, the Hubbard model is solvable
with Bethe-ansatz\cite{lieb}. The system exhibits an interesting
$SO(4)$ symmetry\cite{yang}. One particular feature is that at half-filling,
any small positive on-site interaction would make the system
a Mott-insulator\cite{lieb}. At less than half filling, the low lying
excitations of the system are characterized by the 
universality class of one dimensional interacting electron systems,
the Luttinger liquid. Anderson has suggested that the normal state 
of the high temperature superconductor may be Luttinger-liquid-like,
and various studies have been carried out to investigate spin
and charge dynamics of the Hubbard model and t-J model\cite{yu,mattis}. 

On the other hand, within the context of problems of interacting
electrons, one very interesting subject is the theory of itinerant
magnetism as a consequence of competition of the kinetic energy
and the electron-electron interaction, such as the ferromagnetism 
in the one-hole case in the strong interaction limit of the Hubbard model
and so on\cite{thouless,nag,mattis2,tian,wang2}. 
Recently, there are two conjectures on the Hubbard model
on the complete graph, based on some empirical rules from small 
size diagonalization. The first was concerned with the ground states
of the system at less than or equal to half-filling, the second
was about the case of filling numbers greater than half-filling\cite{salerno}.
It was conjectured that the ground states of the Hubbard model
on complete graph are ferromagnetic when the electron number
$N_e$ is greater than half-filling. In particular, 
it was conjectured that the ground state is unique when
$N_e=L+1$. In this work, I provide 
a proof for this conjecture. Although it is extremely
simple in my opinions, the details of the proof are still presented
carefully, to clarify the unclear points existing in previous literatures.

The Hubbard model defined on a complete graph is given by:
\begin{equation}
H=-|t| \sum_{\sigma, 1\le i\ne j\le L} c_{i\sigma}^\dagger c_{j\sigma}
+ U \sum_{i=1}^L n_{i\downarrow} n_{i\uparrow},
\end{equation}
where the size of the lattice is $L$, the electron number operators
are $n_{i\uparrow}=c_{i\uparrow}^\dagger c_{i\uparrow}, 
n_{i\downarrow}=c_{i\downarrow}^\dagger c_{i\downarrow}$. The electron
creation and annihilation operators obey the usual fermion
anticommutation relations:
$\{c_{i\sigma},c_{j\sigma'}\}=\{c_{i\sigma},c_{j\sigma'}\}=0,
\{c_{i\sigma},c_{j\sigma'}^\dagger\}=\delta_{ij}\delta_{\sigma\sigma'}$.
The on-site energy $U$ is assumed to be positive, $U>0$. 
For finite $U$, the free energy of the system can be obtained 
in the thermodynamic limit $L\rightarrow \infty$\cite{van}, however, 
the wavefunctions of the system were unknown. 
In another limit where the lattice size is finite and $U=\infty$,
the wavefunctions are Gutzwiller projected Fermi sea, and 
they take the Jastrow product form\cite{verg,wang1,wang3}. In the following, we are
interested in the ground state wavefunctions in the most general 
case of finite $L$ and arbitrary but finite $U$.

Assume that the electron number $N_e=\sum_{i=1}^L \sum_{\sigma=\uparrow,\downarrow}
c_{i\sigma}^\dagger c_{i\sigma}=L+1$. Let us work in the Hilbert space
(denoted by $\tilde SH $ ) where the z-component of the total spin 
operator is $S_z=0,(1/2)$ for even (odd) number of electrons on the lattice. 
The lowest energy state in this space would be the ground state of the 
Hamiltonian $H$ in the space of $N_e=L+1$, because the system is 
rotational invariant. In this subspace $\tilde SH$, one
can expand the kinetic operator $T$ in terms of its complete
eigenvectors:
\begin{equation}
T=\sum_{s,K} \sum_{\alpha} \epsilon_{\alpha,s,K}
|\alpha,s,K><\alpha,s,K|,
\end{equation}
where the spin quantum number $s$ is summed from $0$ to $N_e/2$, 
and the quantum number $K$ is the total momentum ( $T$ operator is 
translational invariant ), the Greek letter $\alpha$ stands for all 
other quantum numbers specifying the eigenstates. 
In this space, the potential operator $V$ can also be expanded
in terms of a complete basis of vectors spanning the space:
\begin{eqnarray}
V=&&\sum_{\alpha,\alpha'}\sum_{s}\sum_{K,K'} 
\delta_{K-K',0(mod 2\pi)} V(\alpha,s,K;\alpha',s,K')
|\alpha,s,K><\alpha',s,K'|\nonumber\\
&&=\sum_{A,B} V(A,B) |A><B| \ge U,
\end{eqnarray}
where $V(A,B)=<A|V|B>$. The operator expansion of $V$ can be 
transformed from one complete basis set to another complete
basis set. The last inequality is obtained when using the complete
basis diagonal in sites. 

Suppose that the state vector $(\alpha_g,s_g,K_g)$ is the lowest eigen
state of the operator $T$, with the eigenenergy $\epsilon_0=
\min\{\epsilon_{\alpha,s,K}\}$. If this state vector also satisfies
the following property 
\begin{equation}
V |\alpha_g,s_g,K_g> =U |\alpha_g,s_g,K_g>, 
\label{eq:condition}
\end{equation}
then this state is a ground state of the Hamiltonian
$H=T+V$, with eigenenergy $E_0=\epsilon_0+U$.
The operators $T$ and $V$ for the Hubbard model on a complete graph
satisfy the condition Eq.(\ref{eq:condition}). 
It is easy to analyze the spectrum of the $T$ operator in the 
subspace of fixed total spin $S$ and total momentum $K$. Introduce the Fourier
transformation $c_{k\sigma}^\dagger={1\over \sqrt L}
\sum_{m=1}^L e^{imk} c_{m\sigma}^\dagger$, with
$m=0, {2\pi\over L}, \cdots, {2\pi(L-1)\over L}$. Then the kinetic operator
$T$ becomes 
\begin{equation}
T=-|t| L\sum_{k\sigma} E_k n_{k\sigma} + |t| N_e,
\end{equation}
where $E_k=\delta_{k,0}$ and $n_{k\sigma}$ is the electron 
number operator in $k$ space. In the Hilbert space where 
$S_z=0$ and $s_g=(L-1)/2$, one can construct the following
state vector
\begin{equation}
|G>=(\sum_{i=1}^L c_{i\downarrow}^\dagger c_{i\uparrow})^{L-1\over 2}
(c_{0\uparrow}^\dagger c_{0\downarrow}^\dagger)
\prod_{k={2\pi\over L}}^{2\pi(L-1)\over L} c_{k\uparrow}^\dagger |0>,
\end{equation}
which satisfies the condition Eq.(\ref{eq:condition}). This wavefunction is thus a ground
state of the system in the full Hilbert space $N_e=L+1$. Obviously, 
the ground state has eigenenergy
\begin{equation}
E_0=-|t| 2L + |t| N_e +U,
\end{equation}
where $N_e=L+1$. This ground state is ferromagnetic with spin $s_g=(N_e-2)/2$,
as seen from the way it is constructed.
The way of construction seems to be the only way
that can minimize the kinetic energy and interaction respectively.
If one tries to minimize $H$ using a trial wavefunction
which is a linear combination of the lowest energy state of $T$
operator in the subspace $S=(N_e-4)/2$, the expectation value of $H$
would seem to be greater than $E_0$, because one can have a nonzero chance
to have two pairs of electrons on the lattice. This can be regarded
as a variational argument for the uniqueness of the ground state.  

For the other filling numbers greater than half-filling, similar argument
may be carried out easily. For instance, suppose that the number of 
electrons is $N_e=L+2$. The ground state energy of the system would be 
\begin{equation}
E_0=-2|t| L + |t| N_e + 2U, 
\end{equation}
and the ground state in the Hilbert space of $S_z=0$ would look like
\begin{equation}
|G',k'>=(S^-)^{L-2\over 2} (c_{0\uparrow}^\dagger c_{0\downarrow}^\dagger)
(c_{k'\uparrow}^\dagger c_{k'\downarrow}^\dagger)
\prod_{k(\ne k') {2\pi\over L}}^{2\pi(L-1)\over L} c_{k\uparrow}^\dagger |0>, 
\end{equation}
where $k'$ can be any value in the range $[{2\pi\over L}, {2\pi 2\over L},
\cdots, {2\pi(L-1)\over L}]$.
The ground state wavefunctions all have the total spin $s_g=(N_e-4)/2$,
and ferromagnetism appears in the system. 
For filling numbers $N_e=L+f$, the ground state energy is 
$E_0=-2|t| L + N_e + fU$, and the ground state spin is $s_g=(L-f)/2$.
From the above analysis, one can construct the wavefunctions
explicitly in a similar manner. The rule is pretty straightforward.

One can generalize the simple argument to discuss ground states of 
the system defined on a complete graph, on which the conduction band
interacts with an array of localized impurity spins. The hamiltonian is 
given by
\begin{equation}
H= - |t| \sum_{\sigma=\uparrow,\downarrow} \sum_{1\le i\ne j\le L} 
c_{i\sigma}^\dagger c_{j\sigma}  -|J| \sum_{i=1}^L 
{\vec s_i}\cdot {\vec s_i(f)},
\end{equation}
where at site $i$ there is a localized impurity spin $\vec s_i(f)$ (spin 1/2).
For this impurity lattice system defined on the complete graph, 
the $SO(4)$ symmetry no longer holds, which was overlooked in
previous article\cite{wang1}.  
The electrons interact with the impurity array through ferromagnetic 
exchange. In the Hilbert space where the filling number of the conducting electrons
is just above half-filling, one can find the ground states of the system.
Suppose that $N_e=\sum_{i=1}^L (c_{i\uparrow}^\dagger c_{i\uparrow}
+c_{i\downarrow}^\dagger c_{i\downarrow}) = L+1$, one can construct 
a wavefunction as follows:
\begin{equation}
|G''> =(\sum_{i=1}^L c_{i\downarrow}^\dagger c_{i\uparrow} 
+ \sum_{i=1}^L s_i^-(f) )^{N_e+L-2\over 2} 
(c_{0\uparrow}^\dagger c_{0\downarrow}^\dagger)
\prod_{k={2\pi\over L}}^{2\pi(L-1)\over L} c_{k\uparrow}^\dagger |0>
\bigotimes |\uparrow,\uparrow,\cdots,\uparrow>_f,
\end{equation}
which is the ground state of $H$ in the Hilbert space of $N_e=L+1$, with
eigenenergy given by $E=-|t|2L + N_e |t| -|J| (L-1)/4$.
This ground state is ferromagnetic with total spin $s_g=(N_e+L-2)/2$. 

In summary, in the thermodynamic limit $L\rightarrow \infty$, the free energy
of the Hubbard model defined on the complete graph
consists of two decoupled terms\cite{van}, which is an artifact of the graph.
However, it is still an interesting physical feature that ferromagnetism
appears in this itinerant electron system, at filling numbers just above
the half-filling, because of the electron-electron interaction.
 
I wish to thank D. C. Mattis  for stimulating discussions during his visit
to Lausanne. I also wish to thank C. Gruber, Mo-lin Ge, James T. Liu,
H. Kunz, D. Macris, C. Stafford, X. Q. Wang, X. Zotos and Y. Zhao
for conversations. This work was supported by the Swiss National
Science Foundation.

\end{document}